\documentclass[conference]{IEEEtran}
\usepackage{amsmath, amsthm, amssymb}
\usepackage{amsthm}

\usepackage{subfig}
 \usepackage[dvips]{graphicx}
 \newtheoremstyle{note}
  {3pt}
  {3pt}
  {\itshape}
  {}
  {\itshape}
  {:}
  {.5em}
  {}

\theoremstyle{note}

\newtheorem{theorem}{Theorem}

\newtheorem{definition}{Definition}
\newtheorem{proposition}{Proposition}
\begin{document}
\title{A performance analysis of multi-hop ad hoc networks with adaptive
antenna array systems }

\author{\IEEEauthorblockN{Olfa Ben Sik Ali and Christian
Cardinal} \IEEEauthorblockA{Electrical Engineering Department \\
\'Ecole Polytechnique de Montr\'eal\\ P.O. Box 6079,  Montr\'eal,
Qc, Canada
\\ Email:\{olfa.ben-sik-ali, christian.cardinal\}@polymtl.ca}
\and \IEEEauthorblockN{Fran\c{c}ois Gagnon\\
} \IEEEauthorblockA{Electrical Engineering Department \\ \'Ecole
de Technologie Sup\'erieure\\ 1100, rue Notre-Dame Ouest
Montr\'eal, Qc, Canada \\ Email:francois.gagnon@etsmtl.ca}}
\maketitle
\begin{abstract}
Based on a stochastic geometry framework, we establish an analysis
of the multi-hop spatial reuse aloha protocol (MSR-Aloha) in ad
hoc networks. We compare MSR-Aloha  to a simple routing strategy,
where a node selects the next relay of the treated packet as to be
its nearest receiver with a forward progress toward the final
destination (NFP). In addition, performance gains achieved by
employing adaptive antenna array systems are quantified in this
paper. We derive a tight upper bound on the spatial density of
progress of MSR-Aloha. Our analytical results demonstrate that the
spatial density of progress scales as the square root of the
density of users, and the optimal contention density (that
maximizes the spatial density of progress) is independent of the
density of users. These two facts are consistent with the
observations of Baccelli et al., established through an analytical
lower bound and through simulations in \cite{bac} and \cite{bac2}.
\end{abstract}
\IEEEpeerreviewmaketitle
\section{Introduction}
An ad hoc network is a collection of autonomous nodes that
communicate in a decentralized fashion without relying on a
pre-established infrastructure or on a control unit.
 The
design of communication protocols and  the analysis of their
performance limits in this class of networks have been the subject
of intense investigation over the last decade. This design
involves the definition of strategies and procedures necessary for
transferring data between the nodes of the network, namely, the
development of medium access rules and routing algorithms. The
analysis of the reliability of these protocols in the context of
ad hoc networks is more complex than in the context of cellular or
controlled networks because of the distributed nature of the
former. New lines of research have been introduced using
analytical tools from the theory of stochastic geometry.
Stochastic geometry represents the nodes of the network as
elements of a point process and studies their average behaviour
under predefined communication strategies \cite{stoy,hag}. This
paper follows this methodology and proposes an analytical
evaluation of the performance of two communication protocols under
a random access strategy, namely, the multi-hop spatial reuse
Aloha protocol (MSR-Aloha) \cite{bac2}, and the nearest receiver
with forward progress (NFP) routing \cite{kleip1,sanj}. In
addition, this paper considers the quantification of the
performance improvement that could be reached by employing
adaptive antenna array systems. In point-to-point communication,
multiple antennas can increase the link capacity by providing some
form of
 diversity. In the presence of concurrent transmissions,  the
 signals from multiple antennas can be combined to mitigate
 interference, and consequently, the link reliability is improved \cite{bair,cox}.
 The performance  of multi-antenna systems in the
 context of ad hoc networks has been previously studied
 in \cite{weber6,weber3,asym,olfa3} (and references therein). However, unlike the work
 presented in the current paper, these previous works
  only consider single-hop communication. The main contribution of
 this paper is the derivation of a closed-form expression for the
 expected  progress of packets (i.e., the mean distance covered in one hop by a transmitted packet toward its destination)
  under MSR-Aloha and NFP schemes, and  with nodes
 employing adaptive antenna array systems. The mean progress
 allows to compute the spatial density of progress of the network,
 i.e., the mean distance traversed by all emitted packets per
 unit area toward their intended destinations in a single  time
 slot, which in turn provides direct insight into the network
 transport capacity \cite{bac,gupta}.  The rest of this paper is organized as follows. Section II
describes the system model and the  communication protocols
considered. In section III, we briefly discuss some related works
and position the contribution of this paper. Sections IV and V
present our analytical results as well as some numerical examples.
Finally, section VI concludes the paper.

\section{System model}

\subsection{Network model}
We adopt the so-called stochastic geometric representation of ad
hoc networks, which is described as a planar network formed by a
set of nodes that are randomly located on the points of a
homogeneous Poisson point process (PPP) with density $\lambda$
nodes per unit area. We assume that each node has an infinite
number of packets to transmit, and access the common medium
according to the slotted Aloha protocol, with a predefined
transmission probability $p$. Thus,   by the property of
independent thinning of PPPs \cite{stoy}, the sets of transmitters
and receivers form two independent homogeneous PPPs with density
$\lambda^t=\lambda p$ and $\lambda^r=\lambda(1-p)$, respectively.
\subsection{Channel and capture models}
We assume that every node uses a single transmit antenna  and $L$
receive antennas. A transmitted signal undergoes both large-scale
fading with a path-loss exponent greater than $2$, and small-scale
Rayleigh fading. We assume that the channel coefficients do not
vary during the transmission of one packet. A data packet is said
to be successfully captured by a receiver node if the
signal-to-interference-plus-noise ratio (SINR) perceived by this
node exceeds a prefixed threshold $\beta$.\\
Formally, let us denote  by $\Phi_t=\{X_i, i\in \mathbb{N}\}$ and
$\Phi_r=\{Y_i, i\in \mathbb{N}\}$ the PPPs corresponding to the
transmitter and the receiver sets, respectively; where the
variables $X_i$ and $Y_i$ are random locations of  transmitters
and receivers, respectively. Consider an emitting node  located at
$X_j$ and a receiver node  located at $Y_k$. The signal emitted by
$X_j$ arrives at  $Y_k$ corrupted by interference and noise. Thus
the received signal vector is expressed as:
\begin{equation}
\mathbf{x}=|X_j-Y_k|^{-\alpha/2}\mathbf{h}_{jk} s_j+\sum_{X_i \in
\Phi^t\setminus \{X_j\}}
|X_i-Y_k|^{-\alpha/2}\mathbf{h}_{ik}s_i+\mathbf{n},
\end{equation}
where $s_i$ is the signal emitted by node $X_i$; $\mathbf{h}_{ik}$
is the channel propagation vector between nodes $X_i$ and $Y_k$,
that is  distributed according to the multivariate complex Normal
law with dimension $L$;  $\mathbf{n}$ is a complex Gaussian noise
vector, with variance $\sigma^2$ per dimension; $|X_i-Y_k|$ is the
distance between $X_i$ and $Y_k$; and $\alpha$ is the path-loss
exponent. In statistical antenna array processing, a weight vector
is applied to the received signal, which is chosen based on the
statistics of the data received and optimized under a given
criterion. When the optimum vector, in the sense of SINR
maximization,  is applied, the resulting SINR is \cite{bair,cox}:

\begin{equation}
SINR_{jk}=|X_j-Y_k|^{-\alpha}\mathbf{h}_{jk}^T\mathbf{R}^{-1}\mathbf{h}_{jk},
\end{equation}
where $\mathbf{R}$ is the interference-plus-noise covariance
matrix expressed as: $\mathbf{R}=\sum_{X_i \in \Phi^t\setminus
\{X_j\}}
|X_i-Y_k|^{-\alpha}\mathbf{h}_{ik}\mathbf{h}_{ik}^T+\sigma^2\mathbf{I}_L$,
the operator $^T$ denotes the Hermitian operator, and
$\mathbf{I}_L$ is the identity matrix with dimension $L \times L$.
In our paper \cite{olfa3}, we have established the following key
result on the probability of successful reception:
\begin{proposition} [\cite{olfa3}] Let $X_i$ and $Y_j$ be a transmitter and a receiver
node.  Employing the optimum combining detector, the probability
of successful communication in a Poisson field of interferers and
Rayleigh fading channel is: \small
\begin{equation}
\begin{array}{lcr}
P_s(\lambda^t,d_{ij},\beta, L) =& P(SINR_{ij}\geq \beta)
&~\text{~~~~~~~  ~~~~~~~~~~~~~~~~~~~~ ~} \nonumber
\end{array}
\end{equation}
\begin{equation}
 \label{eqps}\text{~~~~~~~~~~~~} =\sum_{k=0}^{L-1} \frac{( \lambda^t \Delta \beta^{2/\alpha}
d_{ij}^2+\sigma^2 \beta)^k}{k!}   \exp{(-\lambda^t \Delta
\beta^{2/\alpha} d_{ij}^2-\sigma^2 \beta)},
\end{equation}
\normalsize
 where $d_{ij}=|X_i-Y_j|$, $\Delta=2\pi/\alpha
 \Gamma(2/\alpha)\Gamma(1-2/\alpha)$, and $\Gamma$ denotes the
 gamma function.
\end{proposition}
In order to simplify the mathematical analysis,  the noise term
will be ignored in the next sections. This simplification is
reasonable since ad-hoc networks are interference-limited.

\subsection{Communication protocol}

We will now describe the two  communication schemes considered. A
source node has a packet that it wishes to deliver to a distant
destination. Unlike \emph{traditional} routing protocols,  no
specific route (in terms of relays list) is determined in advance.
At each time slot, a source (or a relay) node authorized to
transmit selects the next relay, among the set of  nodes in place,
according to some predefined rules. In NFP routing, a transmitter
selects the next relay to be its closest non-emitting node that
lies  in the direction of the final destination of the packet
processed. Selecting the closest receiver allows  the probability
of successful communication to be maximized. However, this scheme
implies short paths, which means a small spatial reuse factor. In
the MSR-Aloha scheme, the next relay is selected to be the closest
node to the final destination, among the set of receivers that
successfully captured the data packet. This scheme provides the
maximum possible  progress toward the destination in each time
slot, but needs the implementation of an elaborate relay selection
procedure.
\begin{figure}[hpt]
\centering
\includegraphics[width=3.75in, height=2.25in]{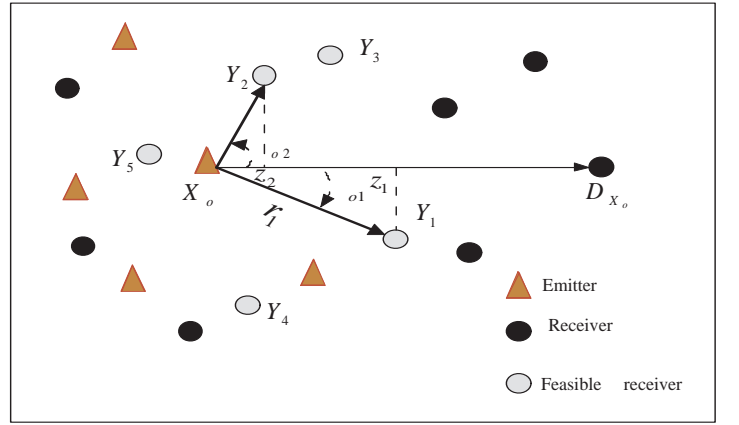}
\caption{Simulation Results} \label{fig1}
\end{figure}
Figure \ref{fig1} shows an example of a snapshot of the network at
an arbitrary time slot. The node $X_o$ is authorized to transmit,
and has a packet for $D_{X_o}$. According to  the NFP scheme,
$Y_2$ is designated to be the next relay, since it is the closest
receiver to $X_o$ in the direction of $D_{X_o}$. Because, in the
case presented by figure \ref{fig1}, $Y_2$ can successfully
capture the signal from $X_o$, the  progress toward $D_{X_o}$,
i.e., the distance travelled by the packet toward $D_{X_o}$, is
equal to $|X_oD_{X_o}-Y_2D_{X_o}|$. This last quantity could be
approximated by $z_2=r_2cos(\theta_{o2})$ if $|X_oD_{X_o}|\gg1$.
Among the set of non-emitting nodes, only $Y_1, Y_2, Y_3, Y_4$,
and $Y_5$ can successfully receive the packet. Applying the
MSR-Aloha scheme, $Y_1$ is the next relay, since it is the closest
to the destination, and thus provides the best progress.
\section{Related works}
\label{work} The idea of designing routing protocols with the
notion of progress is first introduced in \cite{kleip1}, where the
authors propose the most forward within radius (MFR) routing. In
MFR routing, an emitter selects its next relay, among the nodes
within some given range from it,  to be the nearest receiver to
the destination. Baccelli et al. \cite{bac} propose the more
sophisticated selection rule of the MSR-Aloha scheme described
above. However, in their analytical framework, they find this
selection rule to be difficult to manipulate, and so they apply
some modifications  to it. These modifications will be discussed
and compared to our framework in the next sections. In
\cite{webl}, the authors propose and analyze the  longest edge
routing (LER), which applies a similar selection rule as
MSR-Aloha. MSR-Aloha and LER differ in one key aspect that is that
LER does not consider the direction of the intended destination,
which is a challenging analytical aspect. It should be noted that
all  works cited do not consider the use of adaptive antenna array
systems, which is one of the new aspects proposed in this work.
 Finally, practical implementation issues and complete simulation
 packages  of MSR-Aloha routing are considered and  detailed in
 \cite{bac,bac2}.

\section{Performance analysis of NFP and MSR-Aloha schemes}
\subsection{Problem formulation}

Before going through mathematical analyses, we must  formally
define some notions described in the previous section. Let $X_i
\in \Phi_t$ be an arbitrary emitting node.
\begin{definition}[Random set of feasible receivers] The random set
of feasible receivers for $X_i\in \Phi_t$ is formed by the subset
of node in $\Phi_r \cup \{X_i\}$ that can capture the packet of
$X_i$. Denoting this subset by $\mathcal{D}_i$, we have
\cite{bac}:
\begin{equation}
\mathcal{D}_i=\left\{Y_j \in \Phi_r: SINR_{ij} \geq
\beta\right\}\cup \{X_i\}
\end{equation}
\end{definition}
\begin{definition}[Relay selection rules for  NFP and MSR-Aloha] Denoting by $Y^*_{NFP}$ and $Y^*_{MSR}$ the next relays
selected by $X_i$ according to  the NFP and  MSR-Aloha schemes,
respectively, we have:
\begin{equation}
Y^*_{NFP}=\arg \min_{Y_k \in \Phi_r: |\theta_{ik}|\leq \pi/2}
|Y_k-X_i|,
\end{equation}
and
\begin{equation}
\label{eqmfp} Y^*_{MSR}=\arg \max_{Y_k \in \mathcal{D}_i}
cos(\theta_{ik})|Y_k-X_i|,
\end{equation}
where $\theta_{ik}$ is the angle between  the two segments
emerging from $X_i$ and pointing to the direction of its intended
destination and to the direction of  $Y_k$, respectively (Figure
\ref{fig1}). \label{def2}
\end{definition}
Note that, in definition \ref{def2}, the quantity
$cos(\theta_{ik})|Y_k-X_i|$ is an approximation on the progress,
which is very accurate when the final destination is further away
from the transmitter.
\begin{definition}[Spatial density of progress]
The expected progress values for NFP and MSR-Aloha are computed
as: $prog_{NFP}=\mathrm{E}\left[|Y^*_{NFP}| \cos
\theta_{Y^*_{NFP}} P_s(\lambda^t,|Y^*_{NFP}|,\beta)\right]$ and
$prog_{MSR}=\mathrm{E}\left[|Y^*_{MSR}|\right]$, respectively. The
spatial density of progress is simply equal to the transmission
density times the expected progress.
\end{definition}
\subsection{Spatial average of  progress derivation}
\subsubsection{NFP routing}
\begin{theorem} \label{theor2}The expected progress for NFP routing is:
\begin{equation}
prog_{NFP}(\lambda,p)=\frac{1}{\sqrt
\lambda}\sum_{k=0}^{L-1}\frac{\Gamma(k+3/2)(\beta^{2/\alpha}\Delta)^k
p^k(1-p)}{k!((1-p)\pi/2+\Delta \beta^{2/\alpha}p)^{k+3/2}}.
\end{equation}
\end{theorem}
\begin{proof}
The PPPs $\Phi_r$ and $\Phi_t$ are stationary (invariance by
translation and rotation). Thus, we consider without any loss of
generality, a typical emitter located at the center of the network
($X_o=(0,0)$) with an intended destination located on the
horizontal axis. The distribution of the distance separating $X_o$
from its closest receiver in the direction of the destination
verifies \cite{stoy,bac}:
\begin{equation}
P(|Y^*_{NFP}| \leq r)=\exp{(-\frac{\lambda^r \pi}{2 r^2})}.
\end{equation}
Moreover, the angle between the direction of $Y^*_{NFP}$ and the
horizontal axis is uniformly distributed in $[-\pi/2,\pi/2]$.
Consequently, the average progress is:
\begin{equation}
\begin{array}{lcr}
prog_{NFP}(\lambda,p) =& \mathrm{E}_{Y^*_{NFP}}\left[|Y^*_{NFP}|
\cos \theta_{Y^*_{NFP}} P_s(\lambda^t,|Y^*_{NFP}|,\beta)\right]
&~\text{~~~~~~~ ~~~~~~~~~~~~~~~~~~~~ ~} \nonumber
\end{array}
\end{equation}
\begin{equation}
 \label{eqps}\text{~~~~} =\int_{-\frac{\pi}{2}}^{\frac{\pi}{2}} \int_0^\infty r \cos
\theta P(\theta) P(|Y^*_{NFP}|=r) P_s(\lambda^t,r,\beta) dr
d\theta.
\end{equation}
where $\mathrm{E}_x[\cdot]$ denotes the expectation with respect
to the random variable $x$. Replacing $P_s(\lambda^t,r,\beta)$ by
its expression given by relation (\ref{eqps}), we get the result
of  theorem \ref{theor2}.
\end{proof}
\subsubsection{MSR-Aloha routing}
As mentioned in section \ref{work}, the MSR-Aloha scheme was
previously analyzed by Baccelli et al. in \cite{bac}. However,
rather than evaluating the mean progress according to the rule
(\ref{eqmfp}), the authors considered an approximation on it,
which consisted in replacing  the event of successful reception,
i.e., $Y_k \in \mathcal{D}_o$ by its probability of occurrence. In
other words, the following modified rule is used:
\begin{equation}
\label{rm}
 \tilde{Y}^*_{MSR}=\arg \max_{Y_k \in \Phi^r}
P_S(\lambda,d_{ok},\beta) r_k \cos \theta_k.
\end{equation}
The following proposition is demonstrated in \cite{bac}.
\begin{proposition}
The mean progress obtained when using the modified selection rule
(\ref{rm}), denoted as $\widetilde{prog}_{MSR}(\lambda,p)$, is a
lower bound on the mean progress associated to the selection rule
(\ref{eqmfp}), and is expressed as:
\begin{equation}
\label{lower} \widetilde{prog}_{MSR}(\lambda,p)
=\frac{1}{\beta^{1/\alpha}\sqrt{\lambda p 2 \Delta \exp{(1)}}}
\tilde{H}(p,\beta),
\end{equation}
where $\tilde{H}(p,\beta)=\int_0^1 1-\exp{\left(-\frac{1-p}{p}
\frac{G(z)}{2\beta^{2/\alpha}\Delta}\right)}dz$,  and
$G(z)=2\int_{t:\frac{\exp{(t)}}{\sqrt{2\exp{(1)}t}}\leq 1/z}
\arccos{\left(\frac{z\exp{(t)}}{\sqrt{2\exp{(1)}t}}\right)}dt$.
\end{proposition}
The lower bound (\ref{lower}) is difficult to evaluate
numerically. Moreover, it concerns the case of a receiver with a
single antenna, and cannot be easily generalized to the case of
multi-antenna systems. We propose a direct manipulation of the
selection rule (\ref{eqmfp}), and we establish the following
theorem.
\begin{theorem}
\label{themfp} The mean progress of MSR-Aloha  verifies:
\begin{equation}
prog_{MSR}(\lambda,p) \leq \frac{1}{\beta^{1/\alpha}\sqrt{\lambda
p \Delta}} H(p,L,\beta).
\end{equation}
The function  $H(p,L,\beta)$ does not dependent on $\lambda$, and
is expressed as:
\begin{equation}
H(p,L,\beta)=\int_0^\infty 1-\exp{\left(-\frac{1-p}{p}
\frac{F(z)}{\beta^{2/\alpha} \Delta}\right)}dz,
\end{equation}
where $F(z)=\sum_{i=0}^{L-1}\sum_{k=0}^{i}
\frac{\Gamma(1/2+k)\Gamma_{inc}(1/2+i-k,  z^2)}{2 k!(i-k)!}$,  and
$\Gamma_{inc}(k,  z)=\int_z^\infty t^{k-1}\exp{(-t)}dt$ is the
incomplete gamma function.
\end{theorem}
\begin{proof}
Consider again the typical emitter $X_o$ and its intended final
destination placed at the horizontal axis. The event that the
progress, denoted  $Z$,  is less than some value $z$ is equivalent
to the event that all the nodes that capture the packet are
situated in the half-plane $D_z=\{(x,y):x\leq z\}$ ($(x,y)$
denotes the Cartesian coordinates of a point in the plane).  Then
we have:
\begin{equation}
\label{eqind} \mathbf{1}_{(Z<z)}=\prod_{Y_j \in
\Phi_r}(1-\mathbf{1}_{(SINR_{oj}>\beta)}\mathbf{1}_{(Y_j \notin
D_z)}),
\end{equation}
where $\mathbf{1}_{\psi}$ denotes the indicator function, which is
equal to $1$ when the propriety $\psi$ holds, and to $0$
otherwise. From (\ref{eqind}), we have:
\begin{equation}
\label{gen} P(Z<z)=\mathrm{E}_{\Phi_r,\Phi_t}\left[\prod_{Y_j \in
\Phi_r}(1-\mathbf{1}_{(SINR_{oj}>\beta)}\mathbf{1}_{(Y_j \notin
D_z)})\right].
\end{equation}
The right hand side of (\ref{gen}) corresponds to the expression
of a  probability generating functional. The probability
generating functional of a PPP $\Psi(f)=\mathrm{E}_\Phi[\prod_{X
\in \Phi} f(x)]$ is equal to $\exp{(-\lambda \int (1-f(X))dX)}$
\cite{west,webl}. Consequently, the progress distribution
verifies:
\begin{equation}
\label{gen}
P(Z<z)=\mathrm{E}_{\Phi_t}\left[\exp{\left(-\lambda^r\int_{\mathbb{R}^2}
\mathbf{1}_{(SINR_{Y}>\beta)}\mathbf{1}_{(Y \notin D_z)}\right)}
dY\right].
\end{equation}
Applying the Jensen's inequality to the last relation, we get:

\begin{equation}\label{jen}
P(Z<z)\geq \exp{\left(-\lambda^r\int_{\mathbb{R}^2} P_s(\lambda^t,
|Y|^2,\beta) \mathbf{1}_{(Y \notin D_z)}dY\right)}.
\end{equation}
The integral in the right hand side is denoted by $In_L$, and
evaluated as follows:
\small
\begin{equation}
In_L(z)= \sum_{i=0}^{L-1}\int_{ z}^\infty \int_{-\infty}^\infty
\frac{(\gamma(x^2+y^2))^i}{i!} \exp{(-\gamma (x^2+y^2))}dydx.
\end{equation}
\normalsize
 where $\gamma=\lambda^t \Delta \beta^{2/\alpha}$. Substituting $x$ and $y$ by $u=\gamma x^2$
and $v=\gamma y^2$, respectively, we get: \small
\begin{eqnarray}
In_L(z)&=& \sum_{i=0}^{L-1}\sum_{k=0}^{i} \frac{1}{\gamma
k!(i-k)!}\int_{\gamma z^2}^\infty
\frac{1}{2}u^{(i-k)-1/2} \exp{(-u)} du \cdot \nonumber \\
&&\int_{0}^\infty v^{k-1/2} \exp{(-v)}dv
 \\
&=&\sum_{i=0}^{L-1}\sum_{k=0}^{i}
\frac{\Gamma(1/2+k)\Gamma_{inc}(1/2+i-k, \gamma z^2)}{2\gamma
k!(i-k)!}.
\end{eqnarray}
\normalsize
 The mean progress yields to: \small
\begin{equation}
prog_{MSR}(\lambda,p) \leq \int_0^\infty
1-\exp{\left(-\lambda^rIn_L(z)\right)} dz.
\end{equation}
\normalsize
 Applying the substitution $w=\sqrt{\gamma}
z$, we get the result of theorem \ref{themfp}.
\end{proof}
\section{Discussion and numerical results}
Analytical results show that the expected progress depends on the
density $\lambda$ only through the factor $1/\sqrt{\lambda}$.
Thus, the spatial density of progress scales as $\sqrt{\lambda}$,
and the optimal contention density, i.e, the value of $p$ that
maximizes the spatial density of progress, is independent of
$\lambda$. This result is consistent with the scaling law of Gupta
and Kumar \cite{gupta}, and with the observations of Baccelli et
al., made through simulations in \cite{bac2}.
 In the next, we present numerical and simulation
results for NFP and MSR-Aloha. In the simulations, the network
area is set at $10^{6} m^2$ with a density of nodes
$\lambda=10^{-3}$. The experimental mean progress is obtained over
$1000$ independent realizations of the network and results are
normalized to a density equal to $1$.
\begin{figure}[h] \centering
\includegraphics[width=3.75in, height=2.25in]{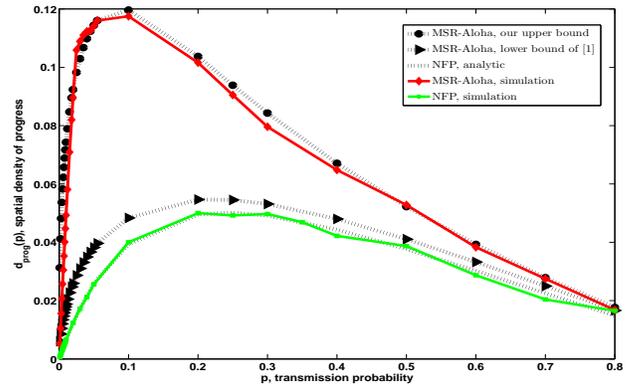}
\caption{Simulation and analytical Results for MSR-Aloha and NFP
routing, with $\alpha=4$, $\beta=1$,  $\lambda=1$ and $L=1$.}
\label{fig2}
\end{figure}

Figure \ref{fig2} presents the spatial density of progress as a
function of the transmission probability, when the number of
receive antennas is set at $1$.   In the same figure, the lower
bound on  the MSR-Aloha spatial density of  progress given by
\cite{bac} is plotted.
 We
observe that our analytical results are very accurate. The lower
bound of \cite{bac} is indeed closer to the curve of NFP routing
than to the curve of  MSR-Aloha scheme, which is a result of the
approximation considered in \cite{bac} that is explained as
follows. The probability of capture decays exponentially with the
distance, and consequently,  the maximization of the product of
the probability of success and the distance (selection rule
(\ref{rm})) is almost equivalent to taking the nearest receiver to
the transmitter, i.e., the NFP scheme.
\begin{figure}[h] \centering
\includegraphics[width=3.75in, height=2.25in]{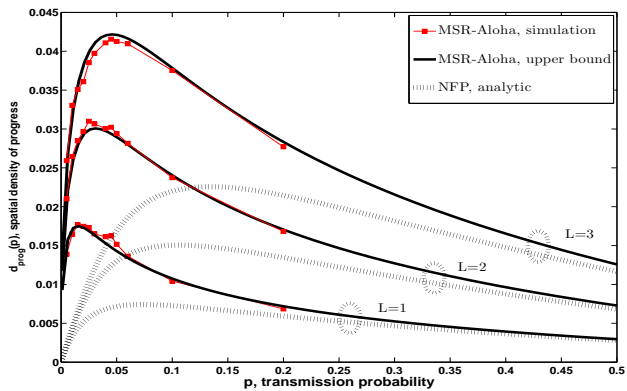}
\caption{Spatial density of progress as a function of the
transmission probability and the number of receive antennas, with
$\alpha=3$, $\beta=10$, and $\lambda=1$. For some value of $p$ the
experimental curves are above the upper bound. This fact is due to
edge effects.} \label{fig21}
\end{figure}
\begin{figure}[h] \centering
\includegraphics[width=3.75in, height=2.25in]{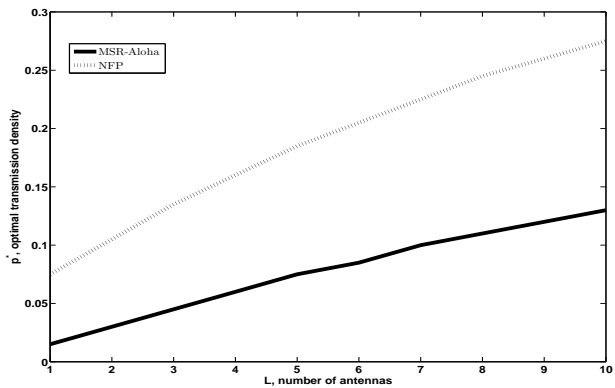}
\caption{ Optimal transmission density as a function of  the
number of receive antennas with $\alpha=3$, $\beta=10$, and
$\lambda=1$.} \label{fig22}
\end{figure}
Figure \ref{fig21} shows that  employing adaptive antenna array
systems significantly improves the spatial density of progress.
For example, with only one additional antenna, a $200\%$ gain is
observed. This improvement is due to the additional diversity and
interference cancellation capability provided by multi-antenna
systems. Observe that the MSR-Aloha scheme is more sensitive to
the transmission probability $p$ than is NFP routing, and this can
be explained as follows: when  $p$ increases, the interference
becomes  severe, leading to a decrease in the distance over which
packets can be captured, and thus, to a significant decrease in
$prog_{MSR}$. For the NFP scheme, the increase in $p$ produces
both good and bad effects. In fact, a low value of $p$ means a
high receiver density, and thus, a high number of receivers in the
vicinity of each transmitter. Consequently, the selection of the
nearest receiver results in  a small amount of progress. On the
other hand, increasing $p$ leads both to an increase in  the
distance separating each transmitter from its nearest receiver and
to a reduction in the probability of successful reception, which
explains the relatively slow variation of the NFP curve as
compared to the MSR-Aloha curve.
\\
Figure \ref{fig22} presents the optimal contention probability as
a function of the number of antennas. For the MSR-Aloha scheme,
the optimal contention probability is equal to $0.015$ for
single-antenna systems (with the parameters indicated on the
figure, the same value was identified by means of simulations in
\cite{bac2}), and to $0.03$ and $0.045$ when using $2$ and $3$
antennas, respectively. Although  NFP allows a higher optimal
contention density, MSR-Aloha has a higher efficiency since it
provides a better progress with a smaller number of transmission
attempts.\\
Finally, several other simulations were done that show the
influence of the parameters $\beta$ and $\alpha$, and these
simulations are not presented here due to a lack of space.

\section{Conclusion}

This paper derived  simple closed-form expressions for the mean
density of progress of two communication strategies, namely
MSR-Aloha and NFP routing. Our results quantify the improvement
achieved through the use of adaptive antenna array systems.
Analytical and simulation results  show that MSR-Aloha protocol is
highly efficient in terms of spatial density of progress and
optimal contention probability. However, this scheme calls for
implementation of a sophisticated relay selection procedure, which
 may introduce additional overhead. A fairer performance
comparison of MSR-Aloha and other communication protocols must
therefore take this aspect into consideration.

\bibliographystyle{IEEEtran}
\bibliography{IEEEabrv,article}

\begin{thebibliography}{10}
\providecommand{\url}[1]{#1}
\csname url@rmstyle\endcsname
\providecommand{\newblock}{\relax}
\providecommand{\bibinfo}[2]{#2}
\providecommand\BIBentrySTDinterwordspacing{\spaceskip=0pt\relax}
\providecommand\BIBentryALTinterwordstretchfactor{4}
\providecommand\BIBentryALTinterwordspacing{\spaceskip=\fontdimen2\font plus
\BIBentryALTinterwordstretchfactor\fontdimen3\font minus
  \fontdimen4\font\relax}
\providecommand\BIBforeignlanguage[2]{{%
\expandafter\ifx\csname l@#1\endcsname\relax
\typeout{** WARNING: IEEEtran.bst: No hyphenation pattern has been}%
\typeout{** loaded for the language `#1'. Using the pattern for}%
\typeout{** the default language instead.}%
\else
\language=\csname l@#1\endcsname
\fi
#2}}

\bibitem{bac}
F.~Baccelli, B.~Blaszczyszyn, and P.~Muhlethaler, ``An aloha protocol for
  multihop mobile wireless networks,'' \emph{{IEEE} Trans. Inf. Theory},
  vol.~52, no.~2, pp. 421--436, 2006.

\bibitem{bac2}
------, ``Time-space opportunistic routing in wireless ad hoc networks:
  Algorithms and performance optimization by stochastic, geometry,'' \emph{the
  {C}omputer Journal}, 2009, to appear.

\bibitem{stoy}
D.~Stoyan, W.~Kendall, and J.~Mecke, \emph{Stochastic Geometry and its
  Application}, ser. Probability and Mathematical Statistics.\hskip 1em plus
  0.5em minus 0.4em\relax Wiley, 1987.

\bibitem{hag}
M.~Haenggi, J.~G. Andrews, F.~Baccelli, O.~Dousse, and M.~Franceschetti,
  ``Stochastic geometry and random graphs for the analysis and design of
  wireless networks,'' \emph{{IEEE} J. Sel. Areas Commun.}, vol.~27, no.~7, pp.
  1029--1046, 2009.

\bibitem{kleip1}
H.~Takagi and L.~Kleinrock, ``Optimal tarnsmission ranges for randomly
  distributed packet radio terminals,'' \emph{{IEEE} Trans. Wireless Commun.},
  vol.~22, no.~3, pp. 246--257, 1984.

\bibitem{sanj}
S.~Biswas and R.~Morris, ``{ExOR}: opportunistic multi-hop routing for wireless
  networks,'' \emph{{ACM SIGCOMM} Computer Communication Review}, vol.~35,
  no.~4, pp. 133--144, 2005.

\bibitem{bair}
C.~A. Baird and C.~Zham, ``Performance criteria for narrowband array
  processing,'' in \emph{{IEEE} Conference On Decision And Control}, vol.~10,
  1971, pp. 564--565.

\bibitem{cox}
H.~Cox, R.~M. Zeskind, and M.~M. Owen, ``Robust adaptive beamforming,''
  \emph{{IEEE} Trans. Acoust., Speech, Signal Process.}, vol.~35, pp.
  1365--1375, 1987.

\bibitem{weber6}
N.~Jindal, S.~P. Weber, and J.~Andrews, ``Rethinking {MIMO} for wireless
  networks: Linear throughput increases with multiple receiver antenna,'' in
  \emph{IEEE {ICC}'09}, 2009.

\bibitem{weber3}
A.~M. Hunter, J.~Andrews, and S.~Weber, ``Transmission capacity of ad hoc
  networks with spatial diversity,'' \emph{{IEEE} Trans. Wireless Commun.},
  vol.~7, no.~12, pp. 5058--5071, 2008.

\bibitem{asym}
S.~Govindasamy, D.~W. Bliss, and D.~H. Staelin, ``Spectral efficiency in
  single-hop ad hoc wireless networks interference using adaptive antenna
  arrays,'' \emph{{IEEE} J. Sel. Areas Commun.}, vol.~25, no.~7, pp.
  1358--1369, 2007.

\bibitem{olfa3}
O.~Ben-Sik-Ali, C.~Cardinal, and F.~Gagnon, ``Performance of optimum combining
  in a poisson field of interferers and rayleigh fading channels,''
  \emph{{IEEE} Trans. Wireless Commun.}, submitted, minor revision, available
  at http://arxiv.org/abs/1001.1482.

\bibitem{gupta}
P.~Gupta and P.~Kumar, ``The capacity of wireless networks,'' \emph{{IEEE}
  Trans. Inf. Theory}, vol.~46, no.~2, pp. 388--404, 2000.

\bibitem{webl}
S.~Weber, N.~Jindal, R.~Ganti, and M.Haenggi, ``Longest edge routing on the
  spatial aloha graph,'' in \emph{{IEEE GLOBECOM'08}}, December 2008.

\bibitem{west}
M.~Westcott, ``The probability generating functional,'' \emph{Journal of the
  Australian Mathematical Society}, vol.~14, pp. 448--466, 1972.

\end{thebibliography}

\newpage

\end{document}